\documentclass[draft,12pt,leqno]{amsart}
\usepackage{amsmath,amsthm,amscd,amssymb}
\usepackage{latexsym}
\begin{document}

\newtheorem{thm}{Theorem}
\newtheorem{cor}[thm]{Corollary}
\newtheorem{lem}[thm]{Lemma}
\newtheorem{prop}[thm]{Proposition}

\theoremstyle{remark}
\newtheorem*{prf}{Proof of Lemma~\ref{lem1}}

\newcommand{\ess}{\operatorname{ess}}
\newcommand{\im}{\operatorname{Im}}
\newcommand{\re}{\operatorname{Re}}
\newcommand{\tr}{\operatorname{tr}}
\newcommand{\sgn}{\operatorname{sgn}}

\title[Spectral Mapping Theorems and 
Invariant Manifolds]{A Spectral Mapping Theorem and 
Invariant Manifolds for Nonlinear Schr\"odinger Equations}

\author[Gesztesy, Jones, Latushkin, and Stanislavova]
{F.~Gesztesy, C.~K.~R.~T.~Jones, Y.~Latushkin, and
M.~Stanislavova}
\address{Department of Mathematics,
University of
Missouri, Columbia, MO
65211, USA}
\email{fritz@math.missouri.edu\newline
\indent{\it URL:}
http://www.math.missouri.edu/people/fgesztesy.html}
\address{Division of Applied Mathematics, Brown University,
Providence, RI 02912, USA}
\email{ckrtj@cfm.brown.edu}
\address{Department of Mathematics, University of
Missouri, Columbia, MO 65211, USA}
\email{yuri@math.missouri.edu}
\address{Department of Mathematics, University of
Missouri, Columbia, MO 65211, USA}
\email{mstanis@pascal.math.missouri.edu}

\begin{abstract} A spectral mapping theorem is proved 
that 
resolves a key problem in applying invariant manifold 
theorems to 
nonlinear Schr\" odinger type equations. The theorem 
is applied to 
the operator that arises as the linearization of the 
equation around 
a standing wave solution. We cast the problem in the 
context of 
space-dependent nonlinearities that arise in optical 
waveguide problems. 
The result is, however, more generally applicable 
including to 
equations in higher dimensions and even systems. The 
consequence 
is that stable, unstable, and center manifolds exist in 
the neighborhood 
of a (stable or unstable) standing wave, such as a 
waveguide mode, 
under simple and commonly verifiable spectral 
conditions.
\end{abstract}
%%\subjclass{Primary: 35Q55, 47D03; Secondary: 47D06}
%%\keywords{Nonlinear Schr\"odinger equation, center manifolds,
%%semigroups, spectral mapping theorem}
\maketitle

%%%%%%%%%%%%%%%%%%%%%%%%%%%%%%%%%%%%%%%%%%%%%%%%%%%%%%%%%%%%
\section{Main Results}
%%%%%%%%%%%%%%%%%%%%%%%%%%%%%%%%%%%%%%%%%%%%%%%%%%%%%%%%%%%%

The local behavior near some distinguished solution, 
such as a steady state, of an evolution equation, can be 
determined through a decomposition into invariant 
manifolds, 
that is, stable, unstable and center manifolds. These 
(locally invariant) 
manifolds are characterized by decay estimates. While 
the flows on the 
stable and unstable manifolds are determined by exponential 
decay in 
forward and backward time respectively, that on the 
center manifold is 
ambiguous. Nevertheless, a determination of the flow on 
the center 
manifold can lead to a complete characterization of the 
local flow and 
thus this decomposition, 
when possible, leads to a reduction of this problem to 
one of identifying 
the flow on the center manifold.

This strategy has a long history for studying the local 
behavior near 
a critical point of an ordinary differential equation, 
or a fixed point 
of a map, and it has gained momentum in the last few 
decades in the 
context of nonlinear wave solutions of evolutionary 
partial differential 
equations. Extending the ideas to partial differential 
equations has, 
however, introduced a number of new issues. In infinite 
dimensions, 
the relation between the linearization and the full 
nonlinear equations 
is more delicate. This issue, however, turns out to be 
not so difficult 
for the invariant manifold decomposition and has largely 
been resolved, 
see, for instance, \cite{Ball}, \cite{BJ}.
A more subtle issue arises at the linear level. All of 
the known proofs 
for the existence of invariant manifolds are based upon 
the use of the 
group (or semigroup) generated by the linearization. 
The hypotheses of 
the relevant theorems are then formulated in terms of 
estimates on the 
appropriate projections of these groups onto stable, 
unstable and center 
subspaces. These amount to spectral estimates that come 
directly from a 
determination of the spectrum of the group. However, in 
any actual 
problem, the information available will, at best, be of 
the spectrum of 
the infinitesimal generator, that is, the linearized 
equation and not its 
solution operator. Relating the spectrum of the 
infinitesimal generator 
to that of the group is a spectral mapping problem 
that is often 
non-trivial. 

In this paper, we resolve this issue for 
 nonlinear Schr\"odinger equations. We formulate the 
results for the 
 case of space-dependent nonlinearities in arbitrary 
dimensions. 
 This class of equations is motivated by the one space 
dimension 
 case that appears in
the study of optical waveguides, see \cite{WGpap}, and has
attracted the
attention of many authors. In particular, there is 
extensive literature on
the existence and instability of standing waves, see, for 
instance, \cite{Gril88,Gril90,GSS,JonesED,JonesJDE} and 
the references therein. In many
instances the questions of the existence
of standing waves and the structure of the spectrum of the 
linearization
of the nonlinear equation around the standing wave are 
well-understood, see \cite{WGpap}.

In the case considered in this paper, the interesting 
examples are known to have the 
spectrum of their
linearization ${\mathcal A}$ enjoying a disjoint 
decomposition: 
the
essential spectrum is positioned on the imaginary axis, and 
there are several isolated
eigenvalues off the imaginary axis, see 
\cite{Gril88,JonesED,JonesJDE}. 
However, as mentioned above, this spectral information
about the linearization ${\mathcal A}$ is not sufficient 
to guarantee the
existence of invariant manifolds. The general theory gives 
the existence of these manifolds for
a semilinear equation with linear part ${\mathcal A}$
only when the spectrum of the operator $e^{t{\mathcal
A}}$, $t>0$, rather than that of ${\mathcal A}$, admits a 
decomposition into disjoint components. 
It is not {\em a priori} clear that the spectrum of the 
operator 
$e^{t{\mathcal A}}$ is obtained from the spectrum of 
${\mathcal A}$ by 
exponentiation. Indeed, in the present case,
the operator ${\mathcal A}$ does not generate a semigroup
for which this property is known (such as for analytic 
semigroups).
Thus, we prove such a spectral mapping theorem 
(cf.~Theorem~\ref{SMTh}) in
this paper. This spectral mapping theorem is derived 
from a known
abstract result in the theory of strongly continuous 
semigroups of
linear operators (see Theorem~\ref{GGHP}). To apply this 
abstract result
one needs to  prove that the norm of the resolvent of 
${\mathcal A}$ is
bounded along vertical lines in the complex plane. The 
corresponding proof
is based on Lemmas~\ref{lem1} and \ref{lem1prime}. The 
main technical step
in the proof of these lemmas concerns a result about the 
high-energy decay
of the norm of a Birman-Schwinger-type operator
(cf.~Proposition~\ref{maintech}), a well-known device 
borrowed from
quantum mechanics.

We consider the following Schr\"odinger equation with 
space-dependent nonlinearity,
\begin{equation}\label{NLS}
iu_t=\Delta u+f(x,|u|^2)u+\beta u,
\quad u=u(x,t)\in {\Bbb C}, \quad x\in {\Bbb R}^n, \quad
t\geq 0,\quad
\beta\in {\Bbb R},
\end{equation}
where
$\Delta = \sum\limits_{j=1}^n \frac{\partial^2}
{\partial x_j} $
denotes the Laplacian, $n \in {\bf N},$ and $f$ is 
real-valued.
Rewriting \eqref{NLS} in terms of its real and 
imaginary parts, $u=v+iw,$
one obtains
\begin{equation}\begin{split} v_t&=
\Delta w+f(x,v^2+w^2)w+\beta w,\\
w_t&=-\Delta v-f(x,v^2+w^2)v-
\beta v.\label{RINLS}\end{split}\end{equation}
A standing wave of frequency $\beta $ for \eqref{NLS} 
is
a time-independent real-valued solution
$\hat{u}=\hat{u}(x)$ of
\eqref{RINLS}. Suppose the standing wave $\hat{u}$ 
is given 
{\em a priori}.
Consider the linearization of \eqref{RINLS}
around
$\hat{u}$ (recalling $\hat{w}=0$), 
\begin{align*}
p_t&=\Delta q+f(x,\hat{u}^2)q+\beta q,\\
q_t&=-\Delta p-f(x,\hat{u}^2)p-2\partial _2
f (x,\hat{u}^2)\hat{u}^2p-\beta p,
\end{align*}
where $\partial _2 f(x,y)=f_y(x,y)$. Thus, the linearized
stability of the standing wave is determined by
the operator
\begin{equation}\label{defA}
{\mathcal A}=
\left[\begin{array}{cc}0 & -L_R\\ L_I & 0\end{array}\right],
\text{ where } L_R=-\Delta-\beta
+Q_1,\quad L_I=-\Delta-\beta +Q_2,
\end{equation}
 and the potentials $Q_1$ and $Q_2$ are explicitly given
by the formulas
$$ Q_1(x)=-f(x,\hat{u}^2(x)),\quad
Q_2(x)=-f(x,\hat{u}^2(x))-2\partial_2
f (x,\hat{u}^2 (x))\hat{u}^2(x).$$

We impose the following conditions on $f,\beta$ and the
standing wave $\hat{u}$ (see
\cite{Gril88,JonesED,JonesJDE}):
\begin{enumerate}\item[(H1)] $f:{\Bbb R}^{n+1}\to {\Bbb R}$
is $C^3$ and all derivatives are bounded on a set of the
form ${\Bbb R}\times U$, where $U$ is a neighborhood of
$0\in {\Bbb R}^n$;
\item[(H2)] $f(x,0)\to 0$ exponentially as $|x|\to  \infty$;
\item[(H3)] $\beta <0$;
\item[(H4)] $|\hat{u}(x)|\to 0$ exponentially as
$|x|\to  \infty$.\end{enumerate}

As a result, the potentials $Q_1$ and $Q_2$
exponentially decay at infinity. The operator
${\mathcal A}$ is
considered on $L^2({\Bbb R^n})\oplus L^2({\Bbb R^n})$;
the domain ${\mathcal D}\left(-\Delta \right)$
is chosen to be the standard
Sobolev space $H^2({\Bbb R}^n)$, and the domain of
${\mathcal A}$ is then
 $H^2({\Bbb R}^n)\oplus H^2({\Bbb R}^n)$. Note, that
 $\left[\begin{array}{cc} 0 & \Delta\\
-\Delta & 0\end{array}\right]$ generates a strongly
continuous group on $L^2({\Bbb R}^n)\oplus L^2({\Bbb R}^n)$.
Thus, its bounded perturbation ${\mathcal A}$ generates
a strongly continuous group
$\{e^{t{\mathcal A}}\}_{t \in{\mathbb R}}$ as well.

It was proved in \cite[Thm.~3.1]{Gril88} that
$\sigma_{\ess}({\mathcal A} )=\{ i\xi
:\xi \in {\Bbb R},|\xi |\geq -\beta \}$.
In addition, it was proved in
\cite{Gril88,JonesED,JonesJDE} that, under the above hypotheses,
$\sigma ({\mathcal A})\backslash \sigma_{\ess}({\mathcal A})$
consists of finitely many eigenvalues, symmetric with
respect to both coordinate axes.

We prove the following result that
relates the spectrum of the semigroup
$\{e^{t{\mathcal A}}\}_{t \geq 0}$
and the spectrum of its generator.

%%%%%%%%%%%%%%%%%%%%%%%%%%%%%%%%%%%%%%%%%%%%%%%%%%%%%%%%%%%%
\begin{thm} [Spectral Mapping Theorem] \label{SMTh} For 
each $n \in {\bf N}$ one has 
 $$
\sigma (e^{t{\mathcal A}})\backslash \{0\}=
e^{t\sigma ({\mathcal A})} \text{ for all } t>0.
$$
\end{thm}
%%%%%%%%%%%%%%%%%%%%%%%%%%%%%%%%%%%%%%%%%%%%%%%%%%%%%%%%%%%%

See \cite{CL,Nagel,Renardy, vanN} for a discussion of the
spectral mapping theorems for strongly continuous
semigroups and examples where the spectral mapping property
as in Theorem~\ref{SMTh} fails. Since the
spectral mapping theorem always holds for the point spectrum,
Theorem~\ref{SMTh} implies, in particular, that
$\sigma_{\ess} (e^{t{\mathcal A}})\subseteq {\Bbb T}
=\{|z|=1\}$. A spectral mapping theorem was proved in 
\cite{kapsan} 
for a nonlinear Schr\" odinger equation with a specific 
potential 
and in the case $n=1$. Their proof also uses 
Theorem~\ref{GGHP} that 
is the key to our result. To the best of our knowledge 
the work of 
Kapitula and Sandstede \cite{kapsan} was the first to 
use the 
Gearhart-Greiner-Herbst-Pr\" uss Theorem in this context. 

It follows that there will be only finitely many 
eigenvalues of 
$e^{t{\mathcal A}}$ off the unit circle and therefore 
general 
results on the existence of invariant
manifolds for semilinear equations can be invoked 
(see, e.g., \cite{BJ}
and compare also with \cite{JBsmall} and the literature 
cited in \cite[p.~4]{LW97}). 

The (local) stable manifold is defined as the set of 
initial data 
whose solutions stay in the prescribed neighborhood and 
tend to 
$\hat{u}$ exponentially as $t \rightarrow +\infty$. 
The unstable 
manifold is defined analogously but in backward time. 
The center
 manifold is complementary to these two and contains 
solutions with 
 neutral decay behavior (although they can decay, they 
will not do 
 so exponentially). In particular, the center manifold 
contains all 
 solutions that stay in the neighborhood in both forward 
and backward 
 time. For details see, for instance, \cite{BJ}.

Concerning the equations under consideration here, we 
have the 
following main theorem.

%%%%%%%%%%%%%%%%%%%%%%%%%%%%%%%%%%%%%%%%%%%%%%%%%%%%%%%%%%%%
\begin{thm}\label{ECM}  Assuming (H1)-(H4), in a 
neighborhood of 
the standing wave solution $\hat {u}$ of \eqref{NLS} there are 
locally invariant stable, unstable and center manifolds. 
Moreover the stable and unstable manifolds are of (equal) 
finite 
dimension and the center manifold is infinite-dimensional. 
\end{thm}
%%%%%%%%%%%%%%%%%%%%%%%%%%%%%%%%%%%%%%%%%%%%%%%%%%%%%%%%%%%%

All of the examples of standing waves considered in \cite{kup} 
and \cite{WGpap} satisfy the hypotheses given here and thus
enjoy a  local decomposition of the flow by invariant 
manifolds. Some of 
these waveguide modes are stable, while others are unstable. In 
the unstable cases, the above results show that the 
instabilities 
are controlled by finite-dimensional (mostly, just 
one-dimensional) 
unstable manifolds. A natural question is whether the 
waveguide modes 
are stable relative to the flow on the center manifold. 
Such a result 
was shown for the case of nonlinear Klein-Gordon 
equations in 
\cite{JBsmall} using an energy argument. Whether such 
an argument 
will work for nonlinear Schr\" odinger equations is open. 
It is more 
than of academic interest, as stability on the center 
manifold has the 
consequence that the center manifold is unique, 
see \cite{BJ}, and armed 
with such a result, a complete description of the local 
flow can be 
legitimately claimed. Cases of standing waves in higher 
dimensions are 
given in \cite{JonesJDE}. Some of these are unstable 
and the above 
considerations again apply. 

We also wish to stress that the spectral mapping theorem 
developed here 
is not restricted to a single equation. Indeed, the 
results formulated 
here are easily adaptable to systems of nonlinear 
Schr\" odinger 
equations. This is particularly important as such 
systems arise in, 
among other problems, second harmonic generation in 
waveguides and 
wave-division multiplexing in optical fibers. 
%%%%%!!!!!!!!!!!!!!
The case of systems is considered in Section~\ref{Sec4}.

In the next section, we give the basic set-up that 
will be used and 
formulate the necessary lemmas for proving 
Theorem~\ref{SMTh}. The 
proofs are given in Section~\ref{Sec3}. 

%%%%%%%%%%%%%%%%%%%%%%%%%%%%%%%%%%%%%%%%%%%%%%%%%%%%%%%%%%%%
\section{Basic lemmas}
%%%%%%%%%%%%%%%%%%%%%%%%%%%%%%%%%%%%%%%%%%%%%%%%%%%%%%%%%%%%

Fix an $a\in{\Bbb R}\backslash\{0\}$ such that the line 
$\{\xi =a+i\tau :\tau
\in {\Bbb R}\}$ does not intersect $\sigma ({\mathcal
A})$. To prove Theorem~\ref{SMTh}, one needs to show that
$\sigma (e^{t{\mathcal A}})$ does not intersect the circle
with radius $e^{ta}$ centered at the origin. We will use
the following
abstract result known as the Gearhart-Greiner-Herbst-Pr\"uss
theorem. 

%%%%%%%%%%%%%%%%%%%%%%%%%%%%%%%%%%%%%%%%%%%%%%%%%%%%%%%%%%%%
\begin{thm} {\rm (}see, e.g., \cite[p.~95]{Nagel}.{\rm )} 
\label{GGHP} Let ${\mathcal A}$ be a generator of a strongly 
continuous semigroup on a complex  Hilbert space. 
Assume
$$
\sigma ({\mathcal A})\cap \{\xi =a+i\tau :\tau \in
{\Bbb R}\}=\emptyset, \quad a\in{\Bbb R}\backslash\{0\}.
$$
Then $\sigma (e^{t{\mathcal A}})\cap \{|z|=e^{ta}\}=
\emptyset$ for all $t>0$, if and only if the norm of the 
resolvent of
${\mathcal A}$ is bounded along the line $\{\xi =
a+i\tau :\tau \in {\Bbb R}\}$, that is,
\begin{equation}
\sup_{\tau \in {\Bbb R}}\|(a+i\tau
-{\mathcal A})^{-1}\|<\infty.\label{BDRes}\end{equation}
\end{thm}
%%%%%%%%%%%%%%%%%%%%%%%%%%%%%%%%%%%%%%%%%%%%%%%%%%%%%%%%%%%%
 Therefore, to prove Theorem~\ref{SMTh}, one needs to
show that $\|(a+i\tau -{\mathcal A})^{-1}\|$ remains
bounded as $|\tau|\to\infty$ for the operator ${\mathcal A}$ 
defined in \eqref{defA}.

We denote $D=-\Delta -\beta$ and recall that
$\beta <0$ by (H3). Moreover, we have
$$\sigma (D)=\sigma
\left(-\Delta\right)-\beta = [-\beta,\infty).$$
We note, that $D^2$ with domain $H^4({\Bbb R}^n)$ is
a self-adjoint operator. Thus, for $\xi =a+i\tau$ with
$\tau \neq 0$ one has $-\xi ^2\notin \sigma (D^2)$. 
Moreover, we write
\begin{equation}\begin{split}\label{f4.1} \xi  -
{\mathcal A} & =\left(\begin{array}{cc} \xi & L_R\\ -L_I & \xi
\end{array}\right) 
= \left(\begin{array}{cc} \xi & D\\ -D & \xi
\end{array}\right) +\left( \begin{array}{cc} 0 & Q_1\\
-Q_2 & 0
\end{array}\right)\\ & =\left(\begin{array}{cc} \xi
& D\\ -D & \xi \end{array}\right) \left[I +\left(
\begin{array}{cc} \xi & D\\ -D & \xi \end{array}\right)^{-1}
\left(\begin{array}{cc} 0 & Q_1\\ -Q_2 &
0\end{array} \right) \right],\end{split}\end{equation}
where, by a direct computation with operator-valued matrices,
\begin{equation}\label{f4.2} \left(\begin{array}{cc} \xi
& D\\ -D & \xi \end{array}\right)^{-1} =\left(
\begin{array}{cc}\xi [\xi^2+D^2]^{-1} & -[\xi^2+D^2]^{-1}D\\
\left[\xi ^2+D^2\right]^{-1}D & \xi
[\xi^2+D^2]^{-1}\end{array}\right) .\end{equation}

%%%%%%%%%%%%%%%%%%%%%%%%%%%%%%%%%%%%%%%%%%%%%%%%%%%%%%%%%%%%
\begin{lem} \label{lem2} For $\xi =a+i\tau,$ 
$a\in{\Bbb R}\backslash\{0\},$ $\tau\in\Bbb R,$ the norm of 
the operator \eqref{f4.2} remains bounded as $|\tau |\to
\infty$. 
\end{lem}
%%%%%%%%%%%%%%%%%%%%%%%%%%%%%%%%%%%%%%%%%%%%%%%%%%%%%%%%%%%%
 The elementary proof of this lemma is given in the next 
section.

Next, we denote
\begin{equation}\label{f5.1} T(\xi)=\left(\begin{array}{cc}
\xi & D\\ -D & \xi \end{array}\right) ^{-1}
\left(\begin{array}{cc} 0 & Q_1 \\-Q_2 & 0\end{array}\right)
=\left[\begin{array}{cc} [\xi^2+D^2]^{-1}DQ_2 & \xi
[\xi^2+D^2]^{-1}Q_1\\
-\xi [\xi^2+D^2]^{-1}Q_2 & [\xi^2+D^2]^{-1}DQ_1\end{array}
\right].
\end{equation}
The main step in the proof of Theorem~\ref{SMTh} is 
contained in the 
next two lemmas. They imply that the
norm of the operator $(I+T(\xi))^{-1}$  is bounded as 
$|\tau|\to\infty$. 
Assume $Q$ is a real-valued potential
exponentially decaying at infinity and let $\xi =a+i\tau $.

%%%%%%%%%%%%%%%%%%%%%%%%%%%%%%%%%%%%%%%%%%%%%%%%%%%%%%%%%%%%
\begin{lem}\label{lem1} 
\begin{enumerate}
\item[(a)] For $n=1,$ $\|Q[\xi^2+D^2]^{-1}D\| \to 0
$ as $|\tau | \to \infty ;$
\item[(b)] For $n \geq 1,$ $\|\xi [\xi^2+D^2]^{-1}
Q\| \to 0 $ as $|\tau | \to \infty $.
\end{enumerate}
\end{lem}
%%%%%%%%%%%%%%%%%%%%%%%%%%%%%%%%%%%%%%%%%%%%%%%%%%%%%%%%%%%%

%%%%%%%%%%%%%%%%%%%%%%%%%%%%%%%%%%%%%%%%%%%%%%%%%%%%%%%%%%%%
\begin{lem}\label{lem1prime}
 For each $n\ge2$ there exists a $\tau_0>0$ such that
for $|\tau |\geq \tau_0$ the operator
$ I + [\xi^2+D^2]^{-1}DQ $ has a bounded inverse. Moreover,
\begin{equation}\label{BDIn}
\sup_{|\tau|\geq \tau_0}  \|[I+[\xi^2+D^2]^{-1}DQ ]^{-1}\|
< \infty.
\end{equation} 
\end{lem}
%%%%%%%%%%%%%%%%%%%%%%%%%%%%%%%%%%%%%%%%%%%%%%%%%%%%%%%%%%%%
 The proofs of Lemmas~\ref{lem1} and \ref{lem1prime} are 
given in the next
section. We proceed finishing the proof of Theorem~\ref{SMTh}. 

%%%%%%%%%%%%%%%%%%%%%%%%%%%%%%%%%%%%%%%%%%%%%%%%%%%%%%%%%%%%
\noindent {\it Proof of Theorem~\ref{SMTh}.}
In the case $n=1,$
by passing to the adjoint operator of
$\xi[\xi^2+D^2]^{-1}Q_1$ and $[\xi^2+D^2]^{-1}DQ_2$,
parts (a) and (b) of Lemma~\ref{lem1} imply that the norm 
of each of the
four block-operators in the right-hand side of \eqref{f5.1} 
is
strictly less than 1 for $|\tau|$ sufficiently large. 
Thus, $\|T(\xi)\|<1$. By
\eqref{f4.1}, one infers
$$\|(a+i\tau -{\mathcal
A})^{-1}\|=\left\|(I+T(\xi))^{-1}\left(\begin{array}{cc}
\xi & D\\ -D & \xi \end{array}\right)^{-1}\right\| \leq
\frac{1}{1-\|T(\xi)\|}\left\| \left(\begin{array}{cc} 
\xi &D\\
-D & \xi \end{array}\right)^{-1}\right\|.$$
Using Lemma~\ref{lem2}, one obtains \eqref{BDRes} and hence
Theorem~\ref{GGHP} implies the result.

For $n \geq 2$, we write $I+T(\xi)=I+M(\xi)+N(\xi)$, where
\begin{equation*}\begin{split}
M(\xi)&= \left[\begin{array}{cc}[\xi^2+D^2]^{-1}DQ_2 & 0\\
0 & [\xi^2+D^2]^{-1}DQ_1 \end{array} \right] , \\
N(\xi)&= \left[\begin{array}{cc}0 
& \xi [\xi^2+D^2]^{-1}Q_1 \\
\xi [\xi^2+D^2]^{-1}Q_2 & 0 \end{array} \right]. 
\end{split}\end{equation*}
By part (b) of Lemma~\ref{lem1}, one has 
$\|N(\xi)\| \to 0 $ as $|\tau|
\to \infty$.
Lemma~\ref{lem1prime} yields that $I+M(\xi)$ is invertible
for $|\tau| \geq \tau_0,$
with $K:=\sup_{|\tau| \geq \tau_0} \|(I+M(\xi))^{-1}\| 
\leq \infty$.
Choosing $\tau'_0 \geq \tau_0$ such that $K\|N(\xi)\| \leq 1$
for $|\tau| \geq \tau'_0,$ the norm of the operator
$ (I+T(\xi))^{-1}=
[I+(I+M(\xi))^{-1}N(\xi)]^{-1}(I+M(\xi))^{-1} $ is
bounded for $|\tau| \geq \tau'_0$ and, as before, 
Theorem~\ref{GGHP}
implies the result. \hfill $\square$\\
%%%%%%%%%%%%%%%%%%%%%%%%%%%%%%%%%%%%%%%%%%%%%%%%%%%%%%%%%%%%

%%%%%%%%%%%%%%%%%%%%%%%%%%%%%%%%%%%%%%%%%%%%%%%%%%%%%%%%%%%%
\section{Proofs}\label{Sec3}
%%%%%%%%%%%%%%%%%%%%%%%%%%%%%%%%%%%%%%%%%%%%%%%%%%%%%%%%%%%%

In this section we give the proofs of Lemmas~\ref{lem2} --
\ref{lem1prime}.  The proof of Lemma~\ref{lem1} is 
based on a direct estimate of the trace norms.
We will give two proofs of  Lemma~\ref{lem1prime}. 
The first proof is applicable to all $n \geq 2$ 
and uses estimates for the norm of the resolvent on 
weighted spaces of 
$L^2$-functions. The second proof 
works for the cases $n=2$ and $n=3$ and is based on 
explicit estimates for the integral kernel of the 
resolvent of the Laplacian.

The main tool in the proof of parts (a) and (b) of
Lemma~\ref{lem1} is the following well-known 
result.~Denote by
${\mathcal J}_q$ the set of operators
$A\in {\mathcal L}(L^2({\Bbb R}^n))$, $n\ge 1$
 such that $\| A\|_{{\mathcal
J}_q}=(\tr(|A|^q))^{1/q}<\infty$, $q\geq 1,$
where $\tr(\cdot)$ denotes the trace of operators in
$L^2({\Bbb R}^n).$ We recall that $\|A\|\leq
\| A\|_{{\mathcal J}_q}$ for all $q\geq 1$.

%%%%%%%%%%%%%%%%%%%%%%%%%%%%%%%%%%%%%%%%%%%%%%%%%%%%%%%%%%%%
\begin{thm} {\rm (}see, e.g., \cite[Theorem XI.20]{RS}.{\rm )} 
\label{ReedSim} 
Suppose $2\leq q<\infty$ and let
$f,g\in L^q({\Bbb R}^n).$ Then $f(x)g(-i\nabla )\in {\mathcal
J}_q$ and
$$\| f(x)g(-i\nabla )\|_{{\mathcal J}_q}\leq
(2\pi )^{-n/2} \| f\|_{L^q}\| g\|_{L^q}.$$
\end{thm}
%%%%%%%%%%%%%%%%%%%%%%%%%%%%%%%%%%%%%%%%%%%%%%%%%%%%%%%%%%%%
 We will apply Theorem~\ref{ReedSim}
to the exponentially decaying $Q=f\in L^q$, $q>1$ and an
appropriate choice of $g$.

%%%%%%%%%%%%%%%%%%%%%%%%%%%%%%%%%%%%%%%%%%%%%%%%%%%%%%%%%%%%
\vspace*{2mm}
\noindent{\it Proof of Lemma~\ref{lem1}.} 
First, let $g(x)=(|x|^2-\beta )(\xi^2+(|x|^2-\beta )^2)^{-1}$.
Then, for $n\ge 1,$ one has $g(-i\nabla )=[\xi^2+
(-\Delta-\beta)^2]^{-1}(-\Delta-\beta)
=g(-i d/dx)=[\xi^2+D^2]^{-1}D$. For
$r=|x|$ one infers,
\begin{equation*}\begin{split} \|g\|^q_{L^q}&=
\int^\infty_0
r^{n-1}(r^2-\beta )^q [(a^2-\tau ^2+(r^2-\beta )^2)^2
+4a^2\tau^2]^{-q/2}dr\\
&=(2|a\tau )|^{-q}\int^\infty_0r^{n-1}(r^2-\beta )^q
\left[\left(\frac{(r^2-\beta )^2-(\tau^2-a^2)}{2|a\tau
|}\right)^2+1\right]^{-q/2}dr.\end{split}\end{equation*}
The change of variable $y=((r^2-\beta )^2-(\tau^2-a^2))
/(2|a\tau |)$ shows that
$$\|g\|^q_{L^q}=k(\tau ) 
\int^\infty_{\frac{\beta^2}{2|a\tau |}
-\tau '}(y^2+1)^{-q/2} \left( \frac{1}{\tau
'}y+1\right)^{\frac{q-1}{2}}\left[\left(\frac{1}{\tau '}y+
1\right)^{\frac{1}{2}}+\frac{\beta } {(\tau
^2-a^2)^{\frac{1}{2}}}\right]^{\frac{n-2}{2}}dy,$$
where
$$k(\tau )=\frac{1}{4} (2|a\tau |)^{1-q} (\tau ^2-
a^2)^{\frac{q-1}{2}+\frac{n-2}{4}}= O(|\tau
|^{\frac{n-2}{2}}) \text{ as } |\tau |\to \infty$$
and
$$\tau'=
(\tau^2-a^2)/(2|a\tau |)=O(|\tau|) \text{ as } |\tau
|\to \infty.$$

We recall, that $\beta <0$ by (H3). Thus,
\begin{equation}\label{esti}
\|g\|^q_{L^q}\leq c|\tau |^{\frac{n-2}{2}}I(-\tau ',\infty),
\end{equation}
denoting
$$I(a,b)=\int^b_a(y^2+1)^{-q/2}\left(\frac{1}{\tau '}y+
1\right)^{\frac{q-1}{2}+\frac{n-2}{4}}dy.$$
Next, one observes that $|\tau |^{\frac{n-2}{2}}I(\tau ',
\infty)\to 0$ as $|\tau |\to \infty$ for
$q>\max\left\{1,\frac{n}{2}\right\}$. Indeed, for
$y\geq \tau'$ one has
$$ 
|\tau |^{\frac{n-2}{2}}I
(\tau ',\infty )\leq |\tau |^{\frac{n-2}{2}}\int_{\tau '}
^{\infty }(y^2)^{-q/2}\left(2\frac{y}{\tau '}
\right)^{\frac{2q+n-4}{4}}dy
=c|\tau |^{\frac{n-2}{2}}{\tau'}^{1-q}\to 0
\text{ as } |\tau |\to \infty.
$$
To prove part (a) of Lemma~\ref{lem1} for $n=1$ we 
remark that
$|\tau |^{\frac{n-2}{2}}I(-\tau ',\tau ')\to 0$ as
$|\tau |\to \infty$.
We stress, that this assertion does not hold for
$n \geq 2$; that is why one needs  Lemma~\ref{lem1prime}.
 Indeed, assuming $n=1$,
for $y\leq \tau '$ one concludes
$$|\tau |^{-\frac{1}{2}}I(-\tau ',\tau )\leq
2^{(2q-3)/4}|\tau
|^{-\frac{1}{2}}\int^\infty_{-\infty}
(y^2+1)^{-q/2}dy\to 0 \text{ as } |\tau |\to \infty.$$
Therefore, for $n=1$ one infers $\|g\|_{L^q}\to 0$ as
$|\tau |\to \infty$ and, using Theorem~\ref{ReedSim},
$$\|Q[\xi^2+D^2]^{-1}D\|\leq \|Q(x)g(-i\nabla )
\|_{{\mathcal J}_q}\leq c\|Q\|_{L^q}\|g\|_{L^q}
\to 0 \text{ as } |\tau|\to \infty.$$
Thus, part (a) of Lemma~\ref{lem1} is proved.

To prove part (b) of Lemma~\ref{lem1}, let
$g(x)=\xi (\xi^2+(|x|^2-\beta )^2)^{-1}.$
Then, for $n\ge 1$, one obtains $g(-i\nabla )=\xi
[\xi^2+(-\Delta-\beta)^2]^{-1}$.
 For $r=|x|$ one concludes as above,
$$\|g\|^q_{L^q}=
l(\tau )\int^\infty_{\frac{\beta^2}{2|a\tau |}
-\tau '} (y^2+1)^{-q/2}\left(\frac{1}{\tau
'}y+1\right) ^{-\frac{1}{2}} \left[ \left(\frac{1}{\tau
'}y+1\right)^{\frac{1}{2}}+\frac{\beta}{(\tau^2-
q^2)^{\frac{1}{2}}}\right]^{\frac{n-2}{2}}dy,$$
where
$$l(\tau )=\frac{1}{4}(2|a\tau |)^{1-q}(\tau ^2-
a^2)^{\frac{n-2}{4}}=O(|\tau|^{\frac{n-2q}{2}})
\text{ as } |\tau |\to \infty.$$
Since $\beta <0$ one has
$$\|g\|^q_{L^q}\leq c|\tau |^{\frac{n-2q}{2}}J(-\tau ',
\infty ),$$
denoting
$$J(a,b)=\int^b_a(y^2+1)^{-q/2}\left(\frac{1}{\tau '}y+
1\right)^{\frac{n-4}{4}}dy.$$
Next, fix $q>\max \{1,(n+2)/2\}$. We claim that 
$|\tau |^{\frac{n-
2q}{2}}J(\tau ', \infty )\to 0$ as $|\tau |\to
\infty$. Indeed, for $y\geq \tau '$ one concludes,
$$
|\tau |^{\frac{n-2q}{2}}J(\tau ',\infty )
\leq |\tau
|^{\frac{n-2q}{2}}\int^\infty_{\tau '} (y^2)^{-q/2}
\left(2\frac{y}{\tau '}\right)^{\frac{n-4}{4}}dy
=c|\tau |^{\frac{2+n-4q}{2}}\to 0 \text{ as } 
|\tau |\to \infty.
$$
On the other hand, for $y\leq \tau '$ one has
$$|\tau |^{\frac{n-2q}{2}}J(-\tau ',\tau ')
\leq 2^{\frac{n-4}{4}}
|\tau |^{\frac{n-2q}{2}}\int^\infty_{-\infty}
(y^2+1)^{-q/2} dy\to 0 \text{ as } |\tau |\to \infty.$$

Therefore, $\| g\|_{L^q}\to 0$ as $|\tau |\to \infty$ and,
using Theorem~\ref{ReedSim}, one obtains for $n\ge 1,$
$$\|Q\xi [\xi^2+(-\Delta-\beta)^2]^{-1}\|\leq \|Q (x)
g(-i\nabla )\|_{{\mathcal J}_q}\leq c\|Q\|_{L^q}\|g\|_{L^q}
\to 0 \text{ as } |\tau
|\to \infty.$$
Thus, part (b) of Lemma~\ref{lem1} is proved. \hfill  
$\square$
%%%%%%%%%%%%%%%%%%%%%%%%%%%%%%%%%%%%%%%%%%%%%%%%%%%%%%%%%%%%

The proof of Lemma~\ref{lem1prime} is based on the
following proposition. Denote 
\begin{equation}\label{DefQ12}
 |Q|^{1/2}(x)=|Q(x)|^{1/2},\quad
Q^{1/2}(x)= |Q|^{1/2}(x)\sgn [Q(x)],
\end{equation}
 so that
$Q=Q^{1/2} |Q^{1/2}|$.
%%%%%%%%%%%%%%%%%%%%%%%%%%%%%%%%%%%%%%%%%%%%%%%%%%%%%%%%%%%%
\begin{prop}\label{maintech}
Assume $Q$ is a real-valued potential 
exponentially decaying at infinity. Then for $n\ge2$
and $\im (\omega) > 0$ one infers
\begin{equation}\label{T9.3}
\||Q|^{1/2}(-\Delta-\omega^2)^{-1}Q^{1/2}\|_{{\mathcal L} 
(L^2({\Bbb R}^n))} 
\rightarrow 0 \ 
\text{ as } \ |\re (\omega^2)| \rightarrow \infty.
\end{equation}
\end{prop}
%%%%%%%%%%%%%%%%%%%%%%%%%%%%%%%%%%%%%%%%%%%%%%%%%%%%%%%%%%%%
 We postpone the proof of the proposition and proceed 
with the proof of Lemma~\ref{lem1prime}.

%%%%%%%%%%%%%%%%%%%%%%%%%%%%%%%%%%%%%%%%%%%%%%%%%%%%%%%%%%%%
\vspace*{2mm}
\noindent {\em Proof of Lemma~\ref{lem1prime}.} Recall
the following elementary fact: 
If $A$ and $B$ are bounded operators, then $I+AB$ is
invertible provided $I+BA$ is invertible; moreover,
$$ (I+AB)^{-1}= I - A (I+BA)^{-1} B. $$
Thus, using $Q=Q^{1/2}|Q|^{1/2}$ and letting
$ A= [\xi^2+D^2]^{-1} D Q^{1/2}$ and $B=|Q|^{1/2},$
the desired inequality ~\eqref{BDIn} is implied by
the following claim 
\begin{equation}
\label{CLA}
\| |Q|^{1/2}[\xi^2+D^2]^{-1} D Q^{1/2} \| \to 0
 \text{ as } |\tau| \to \infty .
\end{equation}
Indeed, if ~\eqref{CLA} holds, then, for some
$\tau_0 > 0$, one has $\|BA\| < 1$ and
$\sup_{|\tau| \geq \tau_0} \| (I+BA)^{-1} \| < \infty$.
To prove $\sup_{|\tau| \geq \tau_0} \| A \| < \infty$
we use the identity
\begin{equation}
\label{Decom}
[\xi^2+D^2]^{-1} D=\frac{1}{2} [D+i \xi]^{-1}
+ \frac{1}{2} [D-i \xi]^{-1}.
\end{equation}
Since $D$ is self-adjoint,
$$\|[D \pm i \xi]^{-1}\| = 1/|\im (\pm i \xi)| =
1/|a| < \infty.$$
This and $\|B\| = \||Q|^{1/2}\|_\infty < \infty$
implies $\sup_{|\tau| \geq \tau_0} \| (I+AB)^{-1} \|
< \infty$, which is the desired relation~\eqref{BDIn}.

To prove the claim \eqref{CLA} one recalls that
$D=-\Delta-\beta $ and
similarly to \eqref{Decom} one obtains
\begin{equation}\label{16.1}
|Q|^{1/2}[\xi^2+D^2]^{-1} D Q^{1/2}=
\frac{1}{2}|Q|^{1/2}[-\Delta-\omega_1^2]^{-1}Q^{1/2}+
\frac{1}{2}|Q|^{1/2}[-\Delta-\omega_2^2]^{-1}Q^{1/2}.
\end{equation}
Here $\omega_1$ and $\omega_2$ are chosen such that
$\im (\omega_1)>0$,
$\im (\omega_2)>0$ and
$\omega_1^2=\beta -i \xi$, $\omega_2^2=\beta +i \xi$.
Recall that $\xi=a+i\tau$. Thus, if $|\tau|\to\infty$ then
$|\re (\omega_j^2)|\to\infty$ for $j=1,2$.
An application of Proposition~\ref{maintech} to each summand 
in the right-hand side of
\eqref{16.1} then proves the claim \eqref{CLA}.
\hfill $\square$
%%%%%%%%%%%%%%%%%%%%%%%%%%%%%%%%%%%%%%%%%%%%%%%%%%%%%%%%%%%%

We will give two proofs of Proposition~\ref{maintech}.
The first proof  is based on an estimate 
for the norm of the resolvent for the Laplacian acting 
between weighted $L^2$-spaces.
This estimate is given in Lemma~\ref{Thm9} below. In fact, 
Lemma~\ref{Thm9} is just
a minor refinement of Lemma~XIII.8.5 in \cite{RS4}. Results 
of this type go back to Agmon~\cite{Ag75}, Ikebe and
Saito~\cite{IS72}, and others (cf. the discussion in 
\cite[p.~345--347]{RS4}). Much  more
sophisticated results of  this type can be found in the 
work by
Jensen~\cite{AJ} and  the bibliography therein.

For $n \geq 2$ and $s>1/2$ let $\rho_s(x)=(1+|x|^2)^{s/2}$, 
for $x \in {\Bbb R}^n$, and
consider the weighted $L^2$-spaces
$$ 
L^2_s({\Bbb R}^n)=\{ f: \|f\|_s:=
\|\rho_s f\|_{L^2({\Bbb R}^n)} < \infty \}
$$ 
and
$$ 
L^2_{-s}({\Bbb R}^n)=\{ f: \|f\|_{-s}:=
\|\rho_s^{-1} f\|_{L^2({\Bbb R}^n)} < \infty \}.
$$
Note that $L^2_s({\Bbb R}^n) 
\hookrightarrow L^2({\Bbb R}^n) \hookrightarrow 
L^2_{-s}({\Bbb R}^n)$. Also, let
${\mathcal S}({\Bbb R}^n)$  denote the Schwartz class of 
rapidly decaying functions.
%%%%%%%%%%%%%%%%%%%%%%%%%%%%%%%%%%%%%%%%%%%%%%%%%%%%%%%%%%%% 
\begin{lem}\label{Thm9} There exists a constant $d=d(n,s)$ 
depending only on $n$ and $s$,
such that for all $\lambda \in {\Bbb C}$ with 
$\im (\lambda) \neq 0$ and 
$| \re (\lambda) | \geq 1, $
and all $\varphi \in {\mathcal S}({\Bbb R}^n)$ the following 
estimate holds 
\begin{equation}\label{T9.1}
\|\varphi\|_{-s} \leq d |\re (\lambda)|^{-1/2} 
\|(-\Delta-\lambda) \varphi\|_s .
\end{equation}
\end{lem}
%%%%%%%%%%%%%%%%%%%%%%%%%%%%%%%%%%%%%%%%%%%%%%%%%%%%%%%%%%%%
\begin{proof} For completeness, we briefly sketch a 
modification of the proof of Lemma XIII.8.5
in \cite{RS4} to prove inequality \eqref{T9.1}.
An elementary calculation shows that for each
$ \lambda \in {\Bbb C}$ with $\im (\lambda) \neq 0$ the 
expression
$$ \gamma ^{-2} := \inf_{y \in {\Bbb R}} \left[ |y^2-
\lambda|^2 + |y|^2 \right]=
\inf_{y \geq 0} \left[ y^2+(1-2 \re (\lambda))y + 
(\re (\lambda))^2+(\im (\lambda))^2 \right]$$
is equal to $\re (\lambda)+(\im (\lambda))^2-1/4$ for 
$\re (\lambda)>1/2$ and to $|\lambda|^2$
for $\re (\lambda) \leq 1/2$. Therefore, $\gamma 
\leq 2 |\re (\lambda)|^{-1/2}$ for 
$|\re (\lambda)| \geq 1$.

For a positive $\alpha$ to be selected below, let 
$\rho_{1,\alpha}(x)=(1+\alpha |x|^2)^{1/2}$. Arguing 
as in the proof of 
Lemma XIII.8.5 in \cite{RS4} (see the corresponding 
equations (61)--(62) in \cite{RS4}), one
infers $$ \|\rho_{1,\alpha}^{-s} \varphi \|_{L^2} \leq 
\gamma \|\rho_{1,\alpha}^{-s} (-\Delta - \lambda) 
\varphi \|_{L^2}+
\gamma (2s \alpha^{1/2}+1) \sum_{j=1}^n 
\|\rho_{1,\alpha}^{-s} \partial_j \varphi \|_{L^2}
+ \gamma (d \alpha 
+ n \alpha^{1/2} s) \|\rho_{1,\alpha}^{-s} 
\varphi \|_{L^2}, $$
where $d$ depends only on $n$ and $s$. Next, pick 
$\alpha < 1$ 
such that
$ d \alpha +n \alpha^{1/2} s < 1/4$. Note that $\alpha$ 
depends only on $n$ and $s$. 
Moreover, $\gamma (d \alpha +n \alpha^{1/2} s) < 1/2$ 
uniformly for 
$|\re (\lambda) | \geq 1,$ since $\gamma \leq 2 
|\re (\lambda) |^{-1/2}$. Since $s>1/2$, 
one obtains
$$ \frac{1}{2} \| \rho_{1,\alpha}^{-s} \varphi 
\|_{L^2} \leq
\gamma (2s+1) (\|\rho_{1,\alpha}^{-s} (-\Delta - \lambda) 
\varphi \|_{L^2}+
\sum_{j=1}^n \|\rho_{1,\alpha}^{-s} \partial_j 
\varphi \|_{L^2}).$$
This and the inequality $\rho_s^{-s} \leq \rho_{1,\alpha}^{-s} 
\leq \alpha^{-s/2} \rho_s^{-s}$
show that 
$$
  \| \varphi \|_{{-s}}=\| \rho_s^{-s} \varphi \|_{L^2} \leq
\| \rho_{1,\alpha}^{-s} \varphi \|_{L^2} \leq
\gamma (2s+1) \alpha^{-s/2} (\| (-\Delta - \lambda) 
\varphi \|_{{-s}}+
\sum_{j=1}^n \| \partial_j \varphi \|_{{-s}}).
$$
Now \eqref{T9.1} follows from the inequalities 
$\|\cdot\|_{{-s}} \leq \|\cdot\|_{{s}}$ and 
$\|\partial_j \varphi \|_{{-s}} \leq C \|(-\Delta - \lambda) 
\varphi \|_{{s}},$
where $C$ is an absolute constant (see Lemma XIII.8.4 
in \cite{RS4}).
\end{proof}
%%%%%%%%%%%%%%%%%%%%%%%%%%%%%%%%%%%%%%%%%%%%%%%%%%%%%%%%%%%%

If $\im (\lambda) \neq 0$, then the resolvent 
$ (-\Delta-\lambda)^{-1}$ is a bounded operator on
$L^2({\Bbb R}^n)$. Consider its restriction 
$R_{s,-s}(\lambda):=(-\Delta-\lambda)^{-1}
\big|_{L^2_s({\Bbb R}^n)}$
as an operator from  $ L^2_s({\Bbb R}^n)$ to 
$ L^2_{-s}({\Bbb R}^n)$. Inequality \eqref{T9.1}
shows that $R_{s,-s}(\lambda)$ is a bounded operator from
$ L^2_s({\Bbb R}^n)$ to $ L^2_{-s}({\Bbb R}^n)$ and that
\begin{equation}\label{T9.2}
\|R_{s,-s}(\lambda)\|_{{\mathcal L }(L^2_s,L^2_{-s})} 
\leq d \ | \re (\lambda) |^{-1/2} \rightarrow 0   
\text{ as } |\re (\lambda)| \rightarrow \infty.
\end{equation}
This relation  with $\lambda=\omega^2$ will be used in 
the first 
proof of Proposition~\ref{maintech}.\

%%%%%%%%%%%%%%%%%%%%%%%%%%%%%%%%%%%%%%%%%%%%%%%%%%%%%%%%%%%%
\noindent {\it First proof of Proposition~\ref{maintech}.} 
Define the multiplication operators $(M_s f ) (x)= 
\rho_s (x) f(x)$ and $(M_{-s} f ) (x)= \rho_s^{-1} (x) f(x)$.
If $Q$ is a real-valued potential exponentially decaying at 
infinity, then using notation \eqref{DefQ12},
one observes that the operators $M_s Q^{1/2}$ and $|Q|^{1/2} 
M_{-s}^{-1} = |Q|^{1/2} M_s$ 
are bounded operators on $L^2({\Bbb R}^n)$. On the 
other hand, 
$M_s : L^2_s({\Bbb R}^n) \rightarrow L^2({\Bbb R}^n) $ and 
$M_{-s} : L^2_{-s}({\Bbb R}^n) \rightarrow L^2({\Bbb R}^n) $ 
are isometric isomorphisms.
This fact and the identity
$$|Q|^{1/2}(-\Delta-\omega^2)^{-1}Q^{1/2} f = 
\left[|Q|^{1/2} M_{-s}^{-1} \right] 
\left[ M_{-s}R_{s,-s}(\omega^2)M_{s}^{-1} \right] 
\left[M_s Q^{1/2} \right] f $$
for $f \in   L^2({\Bbb R}^n)$, yields the following estimate 
\begin{eqnarray*}
&& \||Q|^{1/2}(-\Delta-\omega^2)^{-1}Q^{1/2}\|_{{\mathcal L} 
(L^2)} \\
&& \quad \leq \||Q|^{1/2} M_{-s}^{-1} \|_{{\mathcal L} (L^2)} 
\| M_{-s}R_{s,-s}(\omega^2)M_{s}^{-1} \|_{{\mathcal L} (L^2)} 
\|M_s Q^{1/2} \|_{{\mathcal L} (L^2)} \\
 && \quad \leq \||Q|^{1/2} M_{-s}^{-1} \|_{{\mathcal L} (L^2)} 
 \| R_{s,-s} (\omega^2) \|_{{\mathcal L} (L^2_s,L^2_{-s})}
\|M_s Q^{1/2} \|_{{\mathcal L} (L^2)}.
\end{eqnarray*}
Using relation \eqref{T9.2} for $\lambda=\omega^2$ one concludes
that \eqref{T9.3} holds.\hfill
$\square$
%%%%%%%%%%%%%%%%%%%%%%%%%%%%%%%%%%%%%%%%%%%%%%%%%%%%%%%%%%%%

The main tool in the second proof of 
Proposition~\ref{maintech} in the case $n=3$ is 
Theorem~I.23 in \cite{Simon}, inspired by previous  results of
Zemach and Klein \cite{ZK58}. Assume $n=3$ and
suppose that
the potential $Q$ satisfies the following (Rollnik) condition
(see \cite[p.~3] {Simon}) 
\begin{equation}\label{RolCon}
\int\int\limits_{{\Bbb R}^6} \frac{|Q(x)| 
|Q(y)|}{|x-y|^2} dx dy
< \infty.\end{equation}
Note that ~\eqref{RolCon} trivially holds for exponentially 
decaying continuous $Q$.
 Consider on
$L^2({\Bbb R}^3)$ the operator
$ {\mathcal R}_{\omega} $ with integral kernel
\begin{equation}\label{defR}
R_\omega (x,y)=|Q|^{1/2}(x) \frac{e^{i\omega |x-y|}}
{4\pi |x-y|} Q^{1/2}(y),
\quad x,y \in {\Bbb R}^3, \, x \neq y, \, \im (\omega) \geq 0.
\end{equation}

%%%%%%%%%%%%%%%%%%%%%%%%%%%%%%%%%%%%%%%%%%%%%%%%%%%%%%%%%%%%
\begin{thm}\label{Th7} {\rm (}\cite[Theorem I.23]{Simon}.{\rm )} 
Assume~\eqref{RolCon}. Then
${\mathcal R}_k,$ $ k\in {\Bbb R}$, is a Hilbert-Schmidt 
operator in
$L^2({\Bbb R}^3)$ with $\lim_{|k| \to \infty}
\|{\mathcal R}_k \| = 0$,
$k \in {\Bbb R}$.
Moreover,
\begin{equation}
\label{trform}
\tr ({\mathcal R}^*_k{\mathcal R}_k{\mathcal R}^*_k
{\mathcal R}_k)\to 0 \text{ as }
 |k| \to \infty, \quad k \in {\Bbb R}.
\end{equation}
\end{thm}
%%%%%%%%%%%%%%%%%%%%%%%%%%%%%%%%%%%%%%%%%%%%%%%%%%%%%%%%%%%%
 Relation~\eqref{trform} follows from a direct
calculation of
$\tr ({\mathcal R}^*_k{\mathcal R}_k{\mathcal R}^*_k
{\mathcal R}_k)$,
resulting in
\[\int\int\int\int_{{\Bbb R}^{12}} e^{-i k (|x-y|-|y-z|
+|z-w|-|w-x|)}
\frac{ |Q(x)| |Q(y)| |Q(w)| 
|Q(z)|}{(4\pi)^4|x-y||y-z||z-w||w-x|}
  dx dy dz dw,\]
see \cite[p.~24]{Simon} (and compare also with \eqref{TRS} 
below),
and from the Riemann-Lebesgue lemma. The first assertion
in Theorem~\ref{Th7} follows
from \eqref{trform} and the estimate $\|{\mathcal R}_k
\|^4 = \|{\mathcal R}^*_k{\mathcal R}_k \|^2 \le 
\|{\mathcal R}^*_k{\mathcal R}_k \|^2_{{\mathcal J}_2}
= \tr ({\mathcal R}^*_k{\mathcal R}_k{\mathcal R}^*_k
{\mathcal R}_k)$.

In the case $n=2$ one
replaces the integral kernel 
${e^{i\omega |x-y|}}/{(4\pi |x-y|)}$
in \eqref{defR} by the Hankel function 
$(i/4)H^{(1)}_0(\omega |x-y|)$
of order zero and first kind (cf.~\cite{FG}). We use the 
following result. 

%%%%%%%%%%%%%%%%%%%%%%%%%%%%%%%%%%%%%%%%%%%%%%%%%%%%%%%%%%%%
\begin{thm} {\rm (}\cite{BDO}, \cite[p.~1449]{Cheney}.{\rm )}
\label{Th8} Assume $Q \in L^2({\Bbb R}^2) 
\cap L^{4/3}({\Bbb R}^2)$. 
Consider on
$L^2({\Bbb R}^2)$ the operator
$ {\mathcal K}_\omega $ with integral kernel
\begin{equation}\label{defK}
K_{\omega}(x,y)=|Q|^{1/2}(x) 
\frac{i}{4}H^{(1)}_0(\omega|x-y|) Q^{1/2}(y),
\quad x,y \in {\Bbb R}^2, \, x \neq y, 
\, \im (\omega) \geq 0, \, \omega\neq 0.
\end{equation}
 Then 
\begin{equation}
\label{2dim}
\|{\mathcal K}_\omega \|_{{\mathcal J}_2} 
\leq c\ |\omega|^{-1/2} 
\|Q\|_{L^{4/3}({\Bbb R}^2)}, 
\quad \im (\omega) \geq 0, \, \omega \neq 0.
\end{equation}
\end{thm}
%%%%%%%%%%%%%%%%%%%%%%%%%%%%%%%%%%%%%%%%%%%%%%%%%%%%%%%%%%%%
 Note that $Q \in L^2({\Bbb R}^2) \cap L^{4/3}({\Bbb R}^2)$ 
for exponentially decaying continuous potentials $Q$.

To establish the connection with the discussion in the  
current paper, we recall that
$(-\Delta - \omega^2)^{-1} $ for $\im (\omega) > 0$ is 
an integral
operator with kernel ${e^{i \omega |x-y|}}/{(4\pi |x-y|)}$ 
in the case $n=3$ and with kernel
$(i/4)H^{(1)}_0(\omega |x-y|)$ in the case $n=2$.

%%%%%%%%%%%%%%%%%%%%%%%%%%%%%%%%%%%%%%%%%%%%%%%%%%%%%%%%%%%%
\vspace*{2mm}
\noindent {\it Second proof of Proposition~\ref{maintech}.}
In the case $n=2$ we use Theorem~\ref{Th8} for the operator
${\mathcal K}_{\omega}:= |Q|^{1/2}
(-\Delta-\omega^2)^{-1}Q^{1/2}$ in $L^2({\Bbb R}^2)$
to conclude that 
$\||Q|^{1/2}(-\Delta-\omega^2)^{-1}Q^{1/2}
\| \to 0$ as $|\re (\omega^2)| \to \infty.$ This proves 
Proposition~\ref{maintech} for $n=2.$

In the case $n=3$ one needs one more calculation.
If $\omega=k + i m$ with $m>0$, $k \in {\Bbb R}$,
 then the integral kernel $S_{\omega} (x,y) $
of the Hilbert-Schmidt operator
${\mathcal S}_{\omega}:= |Q|^{1/2}
(-\Delta-\omega^2)^{-1}Q^{1/2}$ in $L^2({\Bbb R}^3)$
is given by the formula
$$S_{\omega} (x,y) = R_{k} (x,y) e^{-m |x-y|},
\quad x,y\in{\mathbb R}^3,\, x\neq y,$$
where $R_{k} (x,y)$ is defined in \eqref{defR} (with 
$\omega =k,$ $\im(\omega)=0$ in \eqref{defR}).

A direct calculation using $m >0$ then shows
\begin{align}
 \tr({\mathcal S}_{\omega}^*{\mathcal S}_{\omega}
{\mathcal S}_{\omega}^*{\mathcal S}_{\omega}) &=
\int\int\int\int_{{\Bbb R}^{12}}
e^{-i k (|x-y|-|y-z|+|z-w|-|w-x|)}
\frac{ |Q(x)| |Q(y)| |Q(w)| |Q(z)|}{|x-y||y-z||z-w||w-x|}
\times \nonumber\\
&  \quad \times (4\pi)^{-4}
e^{- m (|x-y|+|y-z|+|z-w|+|w-x|)} dx dy dz dw
\leq \tr ({\mathcal R}^*_{k}{\mathcal R}_{k}
{\mathcal R}^*_{k}{\mathcal R}_{k}).
\label{TRS}\end{align}
But then relation \eqref{trform} yields  
$\|{\mathcal S}_{\omega}\|^4 \leq
\tr ({\mathcal R}^*_{k}{\mathcal R}_{k}
{\mathcal R}^*_{k}{\mathcal R}_{k})
\to 0$ as $|k|\to\infty.$ Similarly, 
$\|{\mathcal S}_{\omega}\|^4 \leq
\tr({\mathcal S}_{\omega}^*{\mathcal S}_{\omega}
{\mathcal S}_{\omega}^*{\mathcal S}_{\omega})
\to 0$ as $m\to +\infty$ by the dominated convergence 
theorem and hence $\|{\mathcal S}_{\omega}\|\to 0$ 
 as $|\re (\omega^2)|=|k^2-m^2| \to \infty.$ 
Thus, $\||Q|^{1/2}[-\Delta-\omega^2]^{-1}Q^{1/2}
\| \to 0$ as $|\re (\omega^2)| \to \infty$, and hence 
Proposition~\ref{maintech} is proved. \hfill $\square$
%%%%%%%%%%%%%%%%%%%%%%%%%%%%%%%%%%%%%%%%%%%%%%%%%%%%%%%%%%%%

%%%%%%%%%%%%%%%%%%%%%%%%%%%%%%%%%%%%%%%%%%%%%%%%%%%%%%%%%%%%
\vspace*{2mm}
\noindent {\it Proof of Lemma~\ref{lem2}.} We recall that 
$\xi=a+i \tau,$ $a\in{\Bbb R}\backslash\{0\},$ $\tau\in\Bbb R$. 
Since $D^2$ is self-adjoint,
$$\| \xi (\xi^2+D^2)^{-1}\| \leq \frac{|\xi |}
{|\im (\xi ^2)|}=\frac{(a^2+\tau ^2)^{1/2}}{2|a\tau |}
\leq c
\text{ uniformly with respect to } |\tau|.$$
On the other hand, $[\xi^2+D^2]^{-1}D=h(D)$ with $h$
defined as $h(p)=p(\xi^2+p^2)^{-1}$. Therefore,
\begin{equation}\label{f4.3} \| (\xi^2+D^2)^{-1}D\|=
\max_{p\in \sigma (D)}|h(p)|=\max_{p\geq
-\beta}|p|[(a^2-\tau^2+p^2)^2+4a^2
\tau^2]^{-\frac{1}{2}}.\end{equation}

To verify that the right-hand side of \eqref{f4.3} is 
bounded as
$|\tau |\to \infty$, one considers two cases. First, let
$p^2\geq 2\tau^2$. Since $p^2-\tau^2\geq \frac{1}{2}p^2$,
$$\max_{|p|\geq \sqrt{2}|\tau |}|p|[(a^2-
\tau^2+p^2)^2+4a^2\tau ^2]^{-\frac{1}{2}}\leq
\max_{p\in{\Bbb R}}|p|\left[a^2+
\frac{p^2}{2}\right]^{-\frac{1}{2}}
\leq \sqrt{2}.$$
Second, for $p^2\leq 2\tau^2,$ one has
$$\max_{|p|\leq \sqrt{2}|\tau |}
|p|\left[(a^2-\tau^2+p^2)^2+4a^2\tau^2\right]^{-\frac{1}{2}}
\leq\max_{|p|\leq \sqrt{2}|\tau
|}\frac{1}{2|a|}\frac{|p|}{|\tau |} \leq
\frac{1}{\sqrt{2}|a|}.$$
This completes the proof of Lemma~\ref{lem2}.\hfill $\square$

%%%%%%%%%%%%%%%%%%%%%%%%%%%%%%%%%%%%%%%%%%%%%%%%%%%%%%%%%%%%
\section{Systems of nonlinear Schr\"{o}dinger 
equations}\label{Sec4}
%%%%%%%%%%%%%%%%%%%%%%%%%%%%%%%%%%%%%%%%%%%%%%%%%%%%%%%%%%%%

Systems of nonlinear Schr\" odinger equations arise in many 
applications of nonlinear optics. From the perspective of 
the current
paper, the fundamental issue is the same as for a single 
equation, namely
the existence and stability of standing wave solutions. A 
particularly
interesting example of multiple pulses was studied recently 
by Yew
\cite{yew1, yew2, yew3}. The problem of second harmonic 
generation occurs
in a slab waveguide with a quadratically nonlinear response. 
Yew showed
that multiple pulses could be generated from the base 
pulse through a
resonant homoclinic bifurcation. The details of this 
process are not
pertinent for the current work, only the outcome, which 
is the presence of
multiple pulses that Yew has shown to be unstable due to 
real, positive
eigenvalues. It can be shown easily that the essential 
spectrum for these
pulses remains on the imaginary axis and hence the spectral 
configuration
for the linearization at the multiple pulses is analogous 
to the cases 
 considered above for scalar equations. Since the 
technology of this
paper  can be applied to this system, 
Theorem~\ref{ECM} holds in
this case. In the following we show how the above 
considerations can be
adapted to this case of systems.
  
The equations governing the second harmonic generation 
problem are 
a system of coupled nonlinear Schr\"{o}dinger equations 
of the form
\begin{align}\label{sysEq}
i\frac{\partial w}{\partial t}&+\Delta w-
\theta w+\overline{w}v  = 0, \\
i\sigma\frac{\partial v}{\partial t}&+\Delta v-\alpha v
+\frac12 w^2=0,
\notag\end{align}
where $w=w(t,x)$ and $v=v(t,x)$, $x\in{\Bbb R}^n$, 
$n\ge 1$,
 are complex valued functions, and
$\alpha$, $\sigma$, and $\theta$ are positive parameters. 
For the 
one-dimensional case
 $n=1$ the questions of the existence of standing
waves for \eqref{sysEq} and, as mentioned above, the 
structure of the 
spectrum of the linearization around the
standing waves are well-understood, see \cite{yew2, yew3}.

The linearization at a standing wave 
$\hat{u}=(\varphi,\psi)$ is given by
the operator $\displaystyle{
{\mathcal A}=\left[\begin{array}{cc}0 & -L_R\\ L_I &
0\end{array}\right]}$ that has the same structure as 
in \eqref{defA},
but $L_R$ and $L_I$ are now  $2\times 2$ operator 
matrices defined as
follows,
\begin{equation}\label{sysA}
 L_R:=\left[\begin{array}{cc}-\Delta+\theta-\psi 
& -\varphi\\
 -{\varphi}/{\sigma} & (-\Delta+\alpha)/\sigma 
\end{array}\right],
 \quad
L_I:=\left[\begin{array}{cc}-\Delta+\theta+\psi 
& -\varphi\\
 -{\varphi}/{\sigma} & (-\Delta
+\alpha)/\sigma\end{array}\right], 
\end{equation}
where the functions $\varphi$ and $\psi$ are assumed 
to be
continuous and exponentially decaying at infinity.
\begin{thm}\label{th12} The Spectral Mapping Theorem  
holds
for the group generated on the space 
$[L^2({\Bbb R}^n)]^4$ by the operator
${\mathcal A}$ with $L_R$ and $L_I$ defined in 
\eqref{sysA}.
\end{thm}
\begin{proof} As above, one needs to show that 
$\|(a+i\tau-{\mathcal
A})^{-1}\|$ remains bounded as $|\tau|\to\infty$. 
Denote 
\[D=\left[\begin{array}{cc}D_1&0\\0&D_2\end{array}
\right],\quad
\text{where}\quad
D_1=-\Delta+\theta,\,D_2=
-\frac{1}{\sigma}\Delta+\frac{\alpha}{\sigma},\]
and consider the following  matrix potentials 
exponentially decaying at 
infinity, 
\[Q_1=\left[\begin{array}{cc}-\psi
&-\varphi\\-{\varphi}/{\sigma}&0
\end{array}\right]\quad\text{and}\quad Q_2=
\left[\begin{array}{cc}\psi&-\varphi\\
-{\varphi}/{\sigma}&0
\end{array}\right].\]
With this new notation and $\xi=a+i\tau$ formula 
\eqref{f4.1} still holds.

By transposing the second and third rows and columns 
in the $4\times 4$
matrix
$\displaystyle{\left[\begin{array}{cc}\xi&D\\
-D&\xi\end{array}\right
]}$, we remark that this matrix is similar to the 
block-diagonal matrix
with the blocks 
$$\left[\begin{array}{cc}\xi&D_1\\
-D_1&\xi\end{array}\right
]\quad\text{and}\quad 
\left[\begin{array}{cc}\xi&D_2\\-D_2&\xi\end{array}
\right]$$
on the main diagonal and zero remaining entries. 
Lemma~\ref{lem2},
applied to each of the diagonal blocks, shows that 
the norm of the
operator \eqref{f4.2} in the matrix case 
remains bounded as $|\tau|\to\infty$.

Formula \eqref{f5.1} is also valid in the new notation. 
In particular, 
one infers, 
\[\xi[\xi^2+D^2]^{-1}Q_1=\left[\begin{array}{cc}
-\xi[\xi^2+D_1^2]^{-1}
\psi&-\xi[\xi^2+D_1^2]^{-1}\varphi\\-\xi[\xi^2
+D_2]^{-1}\varphi/\sigma&0
\end{array}\right].\]
By part (b) of Lemma~\ref{lem1}, the norm of each 
entry in the last
matrix decays to zero as $|\tau|\to\infty$. Similarly 
for
$\xi[\xi^2+D^2]^{-1}Q_2$, and, as a result, the norms 
of the off-diagonal
entries of $T(\xi)$, defined as in \eqref{f5.1}, decay 
to zero as
$|\tau|\to\infty$.

To handle the diagonal entries of $T(\xi)$, let 
$Q:=[q_{j,k}]_{j,k=1}^2$
be a matrix-valued function with exponentially decaying 
$q_{ij}$. Note 
that
$$ [\xi^2+D^2]^{-1}DQ=
\left[
\begin{matrix}
\,[\xi^2+D_1^2]^{-1}D_1q_{11} &
[\xi^2+D_1^2]^{-1}D_1q_{12}\, \\
\,{[\xi^2+D_2^2]^{-1}D_2q_{21}} &
[\xi^2+D_2^2]^{-1}D_2q_{22}\,
\end{matrix}
\right]. $$
In the case $n=1$, one can apply part (a) of 
Lemma~\ref{lem1} to each
entry of this matrix. This shows that 
$\|T(\xi)\|\to 0$ as
$|\tau|\to\infty$ and the proof of Theorem~\ref{th12} 
for $n=1$ is finished.

In the case $n\ge 2$ we claim that the conclusion of 
Lemma~\ref{lem1prime}
holds for the matrix-valued case (and, hence, 
Theorem~\ref{th12} 
holds). Indeed, to verify \eqref{BDIn}, one considers 
the polar
decomposition $Q(x)=|Q(x)| U(x)$, where 
$|Q(x)|:=[Q^*(x)Q(x)]^{1/2}$
and $U(x)$ is a partial isometry. To be consistent 
with our previous
notations in \eqref{DefQ12}, we denote 
$|Q|^{1/2}(x)=|Q(x)|^{1/2}$
and $Q^{1/2}(x)=|Q(x)|^{1/2}U(x)$, so that 
$Q=|Q|^{1/2}Q^{1/2}$.
Similarly to the proof of Lemma~\ref{lem1prime} in 
the scalar case, 
we put $A=[\xi^2+D^2]^{-1}DQ^{1/2}$ and $B=|Q|^{1/2}$ 
and observe that
$$ [I+[\xi^2+D^2]^{-1}DQ]^{-1}=I-A(I+BA)^{-1}B.$$
As above, the estimate \eqref{BDIn} in the matrix case 
follows
from the relation
\[ \||Q|^{1/2}[\xi^2+D^2]^{-1}D|Q|^{1/2}\|\to 0
\quad\text{as}\quad
|\tau|\to\infty,\]
(cf. claim \eqref{CLA}). To verify this relation, 
we denote 
$|Q|^{1/2}(x):=[a_{j,k}(x)]_{j,k=1}^2$ and observe 
that the entries
of the matrix operator $|Q|^{1/2}[\xi^2+
D^2]^{-1}D|Q|^{1/2}$ are
scalar operators of the type
$$a_{j,1}[\xi^2+D_1^2]D_1a_{1,k}+
a_{j,2}[\xi^2+D_2^2]^{-1}D_2a_{2,k}, \quad j,k=1,2.$$
But the norm of each summand in these entries tends 
to zero as
$|\tau|\to\infty$. To verify the latter fact one can 
apply identity
\eqref{16.1}, Proposition~\ref{maintech}, and 
Lemma~\ref{Thm9}
for the scalar case, and the result follows.
\end{proof}
%%%%%%%%%%%%%%%%%%%%%%%%%%%%%%%%%%%%%%%%%%%%%%%%%%%%%%%%%%%
\vspace*{3mm}
\noindent {\bf Acknowledgments.}

We are indebted to Arne Jensen for valuable correspondence 
in connection with clarifying the origin of results of
the type of Lemma~\ref{Thm9}.

C. Jones was supported by the
National Science Foundation under grant number DMS-9704906.

%%%%%%%%%%%%%%%%%%%%%%%%%%%%%%%%%%%%%%%%%%%%%%%%%%%%%%%%%%%

%%%%%%%%%%%%%%%%%%%%%%%%%%%%%%%%%%%%%%%%%%%%%%%%%%%%%%%%%%%%


\begin{thebibliography}{xxxx}
%%%%%%%%%%%%%%%%%%%%%%%%%%%%%%%%%%%%%%%%%%%%%%%%%%%%%%%%%%%%



\bibitem{Ag75} S.~Agmon, {\em Spectral properties of 
Schr\"odinger 
operators and scattering theory,}  Ann. Scuola Norm. Sup. 
Pisa Ser.~4,  {\bf 2} (1975) 151--218.

\bibitem{Ball} J.~Ball, {\em Saddle point analysis for an 
ordinary differential equation in a 
Banach space and an application to dynamic buckling 
of a beam,}
 {\em In:} Nonlinear Elasticity, R.W. Dickey (ed.), 
 Academic Press, NY (1973) 93-160.

\bibitem{BJ} P.~Bates and C.~Jones,
{\em Invariant manifolds for semilinear partial 
differential
equations,} Dynamics Reported, {\bf 2} (1989) 1--38.

\bibitem{JBsmall} P.~Bates and C.~Jones,
{\em The solution of the nonlinear Klein-Gordon equation
near a steady state,} {\em In:} Advance topics in the 
theory
of dynamical systems, G. Fusco, M. Iannelli, and
L.~Salvadori (eds.) (1989) 1--9.

\bibitem{BDO} D.~Boll\'e, C.~Danneels and T.~A.~Osborn,
{\em Local and global spectral shift functions in 
${\bf R}^2$,} 
J. Math. Phys., {\bf 30} (1989) 420--432. 

\bibitem{FG} D.~Boll\'e, F.~Gesztesy and C.~Danneels,
{\em Threshold scattering in two dimensions,}  Ann. Inst. 
Henri Poincar\'e, 
{\bf 48} (1988) 175--204.

\bibitem{Cheney} M.~Cheney, {\em Two-dimensional 
scattering: 
The number of bound states from scattering data,}
 J. Math. Phys., {\bf 25} (1984) 1449--1455.

\bibitem{CL} C. Chicone and Y. Latushkin,
{\em Evolution Semigroups in Dynamical Systems 
and Differential Equations,}
Math. Surv. and Monogr. {\bf 70}, Amer. Math. Soc., 
Providence, RI, 
to appear.

\bibitem{Gril88} M.~Grillakis,
{\em Linearized instability for nonlinear Schr\"{o}dinger
and Klein-Gordon equations,}
 Commun. Pure Appl. Math., {\bf 41} (1988) 747--774.

\bibitem{Gril90} M.~Grillakis,
{\em Analysis of the linearization around a critical 
point of 
an infinite
dimensional Hamiltonian system,}
 Commun. Pure Appl. Math., {\bf 43} (1990) 299--333.

\bibitem{GSS} M.~Grillakis, J.~Shatah, and W.~Strauss,
{\em Stability theory of solitary waves in the presence of 
symmetry, I, II,}
 J. Funct. Anal., {\bf 74} (1987) 160--197,
{\bf 94} (1990) 308--348.

\bibitem{IS72} T.~Ikebe and Y.~Saito, 
{\em Limiting absorption method 
and absolute continuity for the Schr\"odinger operator,}
 J. Math. Kyoto Univ., {\bf 12} (1972) 513--542.

\bibitem{AJ} A.~Jensen,
 {\em High energy resolvent estimates for Schr\"{o}dinger 
operators in Besov spaces,}
  J. d'Anal. Math., {\bf 59} (1992) 45--50.

\bibitem{JonesED} C.~Jones,
{\em Instability of standing waves for non-linear 
Schr\"{o}dinger
equations,} Ergod. Th. and Dynam. Sys., {\bf 8} 
(1988) 119--138.

\bibitem{JonesJDE} C.~Jones,
{\em An instability mechanism for radially symmetric 
standing 
waves of a nonlinear Schr\"{o}dinger equation,}
 J. Diff. Eqns., {\bf 71} (1988) 34--62.

\bibitem{kup} C.~Jones, T.~K\" upper and K.~Schaffner, 
{\em Bifurcation of asymmetric solutions in nonlinear 
optical media,} 
preprint (1999)

\bibitem{WGpap} C.~Jones and J.~Moloney,
{\em Instability of standing waves in nonlinear optical 
waveguides,}
 Physics Letters A, {\bf 117} (1986) 175--180.

\bibitem{kapsan} T.~Kapitula and B.~Sandstede, 
{\em Stability of bright solitary wave solutions to 
perturbed 
nonlinear Schr\" odinger equations,}
 Physica D, {\bf 124} (1998) 58-103.

\bibitem{LW97} C.~Li and S.~Wiggins, {\em Invariant 
Manifolds and Fibrations for Perturbed Nonlinear 
Schr\"odinger Equations}, Springer-Verlag, New York, 1997.

\bibitem{Nagel} R.~Nagel (ed.), 
{\em One Parameters Semigroups of Positive Operators,}
Lecture Notes in Math., {\bf 1184}, Springer-Verlag,
Berlin, 1984.

\bibitem{RS} M. Reed and B.~Simon,
{\em Methods of Modern Mathematical Physics. III:
Scattering Theory,} Academic Press, New York, 1979.

\bibitem{RS4} M.~Reed and B.~Simon,
{\em Methods of Modern Mathematical Physics. IV:
Analysis of Operators,} Academic Press, New York, 1978.

\bibitem{Renardy} M.~Renardy, {\em On the
linear stability of hyperbolic PDEs and
viscoelastic flows,} Z. Angew. Math.
Phys., {\bf 45} (1994) 854--865.

\bibitem{Simon}  B.~Simon,
{\em Quantum Mechanics for Hamiltonians Defined as 
Quadratic
Forms,} Princeton Univ. Press, Princeton, 1971.

\bibitem{vanN} J.~M.~A.~M.~van Neerven,
{\em The Asymptotic Behavior of Semigroups of Linear 
Operators,}
Operator Theory. Advances and Applications {\bf 88}, 
Birkh\"{a}user, Basel, 1996. 

\bibitem{yew1} A.~Yew,
{\em Multipulses of nonlinearly coupled Schr\" odinger 
equations},
preprint, 1999.

\bibitem{yew2} A.~Yew, 
{\em Stability analysis of multipulses in nonlinearly 
coupled 
Scr\" odinger equations}, preprint, 1999.

\bibitem{yew3} A.~Yew, 
{\em An analytical study of solitary waves in 
quadratic media},
Ph.D. Thesis, Brown University, Providence, RI, 1998.

\bibitem{ZK58} C.~Zemach and A.~Klein, {\em The Born
expansion in non-relativistic quantum theory,}  Nuovo
Cim., {\bf 10} (1958) 1078--1087.


\end{thebibliography}
\end{document}